\documentclass[aps,twocolumn,showpacs,preprintnumbers,amsmath]{revtex4}

\usepackage{amssymb}
\usepackage{graphicx}
\usepackage{dcolumn}
\usepackage{bm}

\begin{document}

\title{Particles Production in Neutron Stars by Means of the Parametric Resonance Mechanism}

\author{S. D. Campos} \email{sergiodc@ufscar.br}
\affiliation{Universidade Federal de S\~ao Carlos, \textit{campus} de Sorocaba, 18052-780, Sorocaba, SP, Brazil.}

\date{\today}

\begin{abstract}
Using a Lagrangean toy model the coupling of a negative pion superconducting field and the electromagnetic field of the star is analyzed. A numerical study of particles produced in the neutron star medium by means of the well-known parametric resonance phenomenon is performed.   
\end{abstract}

\pacs{97.60.Jd, 26.60.-c}

\maketitle

\section{Introduction}
In quantum field theory the parametric resonance \cite{landau1} describes the resonant amplification of quantum fluctuations, which can be classically viewed as particle production. In such way, it was first used as an effective particle production mechanism, in the Post-Inflationary Cosmology context, by Traschen and Brandenberger \cite{traschen}, Dolgov and Kirilova \cite{dolgov}, and few years latter by Kofman \textit{et al} \cite{kofman}, and may play a fundamental role in the reheating of the universe after the inflationary era. 

In the cosmological context, the production of gravitational waves due to quantum fluctuations of the vacuum during the transition between the inflationary period and the radiation-dominated era may present this phenomenon, but the main features in the spectrum are due to the inflaton field and not due to the resonance field and then the gravitational waves produced are not enough to be seen by the interferometer detectors \cite{sa}. On the other hand, in the preheating phase, this phenomenon leads to large inhomogeneities which source a stochastic background of gravitational waves at scales inside the comoving Hubble horizon \cite{easther}. Investigations of reheating in chaotic inflation and in hybrid inflation models have shown that reheating may occur much faster, due to nonperturbative effects such as parametric resonance and exponential growth of tachyonic modes \cite{linde}. A very interesting investigation of a gauge field coupled with a charged scalar field is done in \cite{finelli} and reveals that, in some conditions, the parametric resonance mechanism could have some relevance to the problem of large scale primordial magnetic fields.

Otherwise, in the context of compact objects, Garc\'{\i}a-Bellido and Kusenko \cite{garcia} had showed that in the merge of two neutron stars the parametric resonance could be obtained from the interaction of the strong electromagnetic field of the resulting star and the superconducting proton field produced in this espectacular event. Photons created due the parametric resonance mechanism could reach the star surface and eventually gamma-ray bursts observed in the Earth sky may be explained by this event. 

Following the Garc\'{\i}a-Bellido and Kusenko's idea, in a previous work \cite{sdc} was proposed that parametric resonance may be obtained in the interior of only one star. In that paper the electromagnetic field of the star interacts with a negative pion condensate leading to a Lam\'e equation type, analytically solved. This equation possess exponential solutions in the resonant bands, classically viewed as particles production. Then, a negative pion condensate field interacting with the strong electromagnetic field of the star may lead to an exponential photoproduction the in the resonant bands.

Here, it is analyzed the coupling of a negative pion superconducting field with the electromagnetic field of the star. Then, this paper is a natural extension of the previous one. This two approaches on the same problem are due to the coupling differences between the electromagnetic field and the negative pion field. If the negative pion field behaves as a condensate, the resulting equations and treatment is presented in \cite{sdc}. If the negative pion field behaves as a superconducting field, then the resulting equations and mathematical treatment is presented here.

In Section II, the pion superconducting phase transition is analyzed.

In Section III, the order parameter presented in the Ginzburg-Landau superconductivity theory \cite{landau2} is used to describe the negative superconducting pion field. This phenomenological approach is valid only near the transition point, i.e., near the critical temperature $T_c$ \cite{landau2}.

In Section IV one propose a Lagrangean formalism including an interaction term between the order parameter and the electromagnetic field of the neutron star. This interaction term can be obtained directly from the Ginzburg-Landau theory \cite{landau2}. A numerical analysis of the particles number and energy density produced is performed.

In Section V are presented the final remarks.     

\section{Superconducting Phase Transition}

Neutron star interior constituents is a matter of discussion, since different approaches allows to different equations of state each one resulting, for example, in different mass ranges, mass-radius relationship and cooling rates \cite{pethick,glendenning}. Here, roughly speaking, one consider a neutron star interior composed of three simple regions defined as outer crust, inner crust and the core, each one with different states of matter. The outer crust consists basically of a lattice of atomic nuclei and Fermi liquid of relativistic, degenerate electrons. The inner crust contains matter in the density range 4$\times$10$^{11}$ g/cm$^3$ (neutron drip) to 2$\times$10$^{14}$ g/cm$^3$ (transition density). The core presents density beyond the transition density, i.e., above the nuclear density $\rho_0\simeq 2.8\times 10^{14}$ g/cm$^3$ and the atomic nuclei have dissolved into their constituents.

The pion condensation is a well-known problem in nuclear matter applied to neutron star interior. Pion condensation in nuclear or neutron matter was initially proposed by Migdal \cite{migdal1}, Sawyer \cite{sawyer} and Scalapino \cite{scalapino} and it is argued that its existence in neutron star medium may affect the thermal evolution of the star, enhance its cooling rate \cite{glendenning}, for example. 

The positive pion field may condensate with a consequent transition to a superconducting state, but Sawyer and Yao \cite{sawyeryao} had showed that the number of positive pion particles is much smaller than the number of negative pion particles. In such way, the possibility of a positive pion superconducting state is negligible.

Neutral pion may condensate in the same way as the charged pion but does not interact with the electromagnetic field of the star. Hence its eventual contributions to the problem of particle production by the parametric resonance are not take in to account in this work. 

Therefore, neglecting strong correlations of pions with the surrounding matter in modifying the pion self-energy, then the decay
\begin{eqnarray}
\nonumber n\rightarrow p+\pi^-,
\end{eqnarray}
\noindent is favorable when $m_{\pi}<\mu_{\pi}=\mu_{n}-\mu_{p}$, where $m_{\pi}=140$ MeV is the $\pi^-$ rest mass, $\mu_{\pi}$, $\mu_{n}$, and $\mu_{p}$ are the chemical potentials of the pion, neutron and proton, respectively. The neutron and proton chemical potential are defined by 
\begin{eqnarray}
\nonumber \mu_n=\frac{{p_F^n}^2}{2M_n}, \hspace{0.3cm} \mu_p=\frac{{p_F^p}^2}{2M_n},
\end{eqnarray}
\noindent where $p_F$ is the usual Fermi momentum calculated for neutrons and protons, and $M_n$ is the neutron mass. The Fermi momenta are related with proton and neutron densities, $\rho_p$ and $\rho_n$, by
\begin{eqnarray}
\nonumber p_F^{(p,n)}=[3\pi^2\rho_{(p,n)}]^{1/3}.
\end{eqnarray}

The total baryon density is $\rho=\rho_p+\rho_n$, then
\begin{eqnarray}
\nonumber \mu_{\pi}=\frac{(3\pi\rho)^{2/3}}{2M_n}\left[\left(1-\frac{\rho_p}{\rho_n}\right)^{2/3}-\left(\frac{\rho_p}{\rho_n}\right)^{2/3}\right],
\end{eqnarray}
\noindent where the ratio $\rho_p/\rho_n$ is determined, in principle, by charge neutrality \cite{baym}.

The electron chemical potential is $\sim$100 MeV at nuclear density. Take into account the interaction with the surrounding matter the pion self-energy may be enhanced and the critical density value for pion condensation may vary from $\rho_0-4\rho_0$. Therefore, the pion condensation may only occur in the neutron star core. 


In the presence of strong electromagnetic field the pion condensate may becomes into a superconducting field \cite{hs1,migdal}. In a phenomenological context, the superconducting field characterization depends on the value of the Ginzburg-Landau parameter $\kappa$ \cite{landau2}. If $\kappa<1/\sqrt{2}$ the superconductor is type-I, and if $\kappa>1/\sqrt{2}$ is type-II and if the condensate is homogeneous, $\kappa$ practically does not depend on $\rho-\rho_c$, where $\rho_c$ is the critical density \cite{migdal}. 

The existence of such superconducting field depends strongly on the magnetic fields values, i.e., there is a critical value $H_c$ that determines its existence and its value is

\begin{eqnarray}
\nonumber H_c=\mu_{\pi}f_{\pi}\left[1-\left(\frac{m_{\pi}}{\mu_{\pi}}\right)^2\right]\left[1+\left(\frac{\mu_{\pi}}{6m_{\pi}}\right)\right]^{1/2}.
\end{eqnarray}

Using the formalism proposed by Harrington and Shepard \cite{hs1}, the upper and lower values of the critical magnetic field are given by
\begin{eqnarray}
\nonumber H_{c2}=H_c\sqrt{2}\kappa, \hspace{0.3cm} H_{c1}=\frac{H_c}{\sqrt{2}\kappa}(\ln{\kappa}+0.08),
\end{eqnarray} 
\noindent where the expression for $H_{c1}$ is only valid in the extreme type-II case. Therefore, only in the range between $H_{c1}$ and $H_{c2}$ one may define the negative pion superconducting field.

The Harrington and Shepard approach is mathematically identical to the Ginzburg-Landau theory of superconductivity and therefore the values obtained for the limiting magnetic fields are valid in the Ginzburg-Landau approach. 

In the mean field approximation the critical temperature is related with the superconducting field, by (Z=N) \cite{migdal}
\begin{eqnarray}
\nonumber T_c^{MF}\approx \epsilon_F\left(\frac{\rho-\rho_c}{\rho_c}\right)^{1/2}, \hspace{0.3cm}T<\epsilon_F,
\end{eqnarray}

For (Z$<$N)
\begin{eqnarray}
\nonumber T_c^{MF}\approx \epsilon_F\left(\frac{\rho_c-\rho}{\rho_c}\right)^{1/2}, \hspace{0.3cm}T<<\epsilon_F,
\end{eqnarray}

Phase fluctuations near the critical temperature must be take into account \cite{migdal}
\begin{eqnarray}
\nonumber T_c\approx \frac{8\pi}{\ln R/z_0},
\end{eqnarray}
\noindent where $z_0\approx R^2$. For a neutron star with $R=$10 Km one obtains $T_c\simeq$10 MeV. 

\section{The Order Parameter}

Considering the superconducting field composed only by negative pion it is possible to suppose that the order parameter, or functional, present in the Ginzburg-Landau theory is sufficient to describe its evolution. Considering a region in the neutron star medium where the Ginzburg-Landau functional may be applied, its equation of motion is written as ($\hbar=c=1$) \cite{landau2} 
\begin{eqnarray}
\label{1}\ddot{\phi}+\frac{8{\epsilon}_F}{3a}\dot{\phi}-\frac{2{\epsilon}_F}{3cm}{\triangledown}^2\phi+U'(\phi)=0,
\end{eqnarray}
\noindent where
\begin{eqnarray}
\nonumber a=\frac{28\zeta(3){\epsilon}_F}{3{\pi}^2T_c},
\end{eqnarray}
\noindent is a constant characterizing the superconducting pion field and
\begin{eqnarray}
\nonumber \epsilon_F=\frac{{p_F}^2}{2m_{\pi}},
\end{eqnarray}
\noindent is the Fermi energy. As usually, $p_F$ is the Fermi momentum, $m_{\pi}$ is the
effective pion mass in nuclear matter, and $\zeta(3)$ is the Riemann Zeta function evaluated in 3 ($\simeq 1.2020569...$). The prime denote differentiation with respect to $\phi$. 

Assuming isotropic distribution of the superconducting field the gradient term may be neglected, then
\begin{eqnarray}
\label{2}\ddot{\phi}+\frac{8{\epsilon}_F}{3c}\dot{\phi}+U'(\phi)=0.
\end{eqnarray}

The coefficient of the dumping term may be written in terms of the critical temperature as
\begin{eqnarray}
\nonumber \frac{8{\epsilon}_F}{3a}=\frac{8\pi^2}{28\zeta(3)}T_c.
\end{eqnarray}

Otherwise, the critical temperature may be expressed in terms of the critical field $H_c$ as (near the critical point) \cite{landau2}
\begin{eqnarray}
\label{3} H_c=2\alpha\sqrt{\frac{\pi}{\rho}}(T_c-T),
\end{eqnarray}
\noindent where $\rho$ is a positive coefficient depending only on the density of the superconducting pion field; $\alpha$ is a positive parameter. Using (\ref{3}) one write (\ref{2}) as 
\begin{eqnarray}
\label{4}\ddot{\phi}-\frac{4\pi^2}{28\alpha\zeta(3)}\sqrt{\frac{\rho}{\pi}}H_c\dot{\phi}+U'(\phi)=0.
\end{eqnarray}

The effective potential $U(\phi)$ may be written as (from the well-known $\lambda\phi^4$ theory)
\begin{eqnarray}
\nonumber U(\phi)={\alpha}(T-T_c)\phi^2+\frac{\rho}{2}\phi^4,
\end{eqnarray}
\noindent or using $H_c$ as
\begin{eqnarray}
\label{5} U(\phi)=-\frac{H_c}{2}\sqrt{\frac{\rho}{\pi}}\phi^2+\frac{\rho}{2}\phi^4,
\end{eqnarray}
\noindent with points of minima
\begin{eqnarray}
\label{6}\phi_0^2=\frac{H_c}{2\rho}\sqrt{\frac{\rho}{\pi}}.
\end{eqnarray}

\begin{figure*}[ht]
\centering
\includegraphics[width=14.0cm,height=10.0cm]{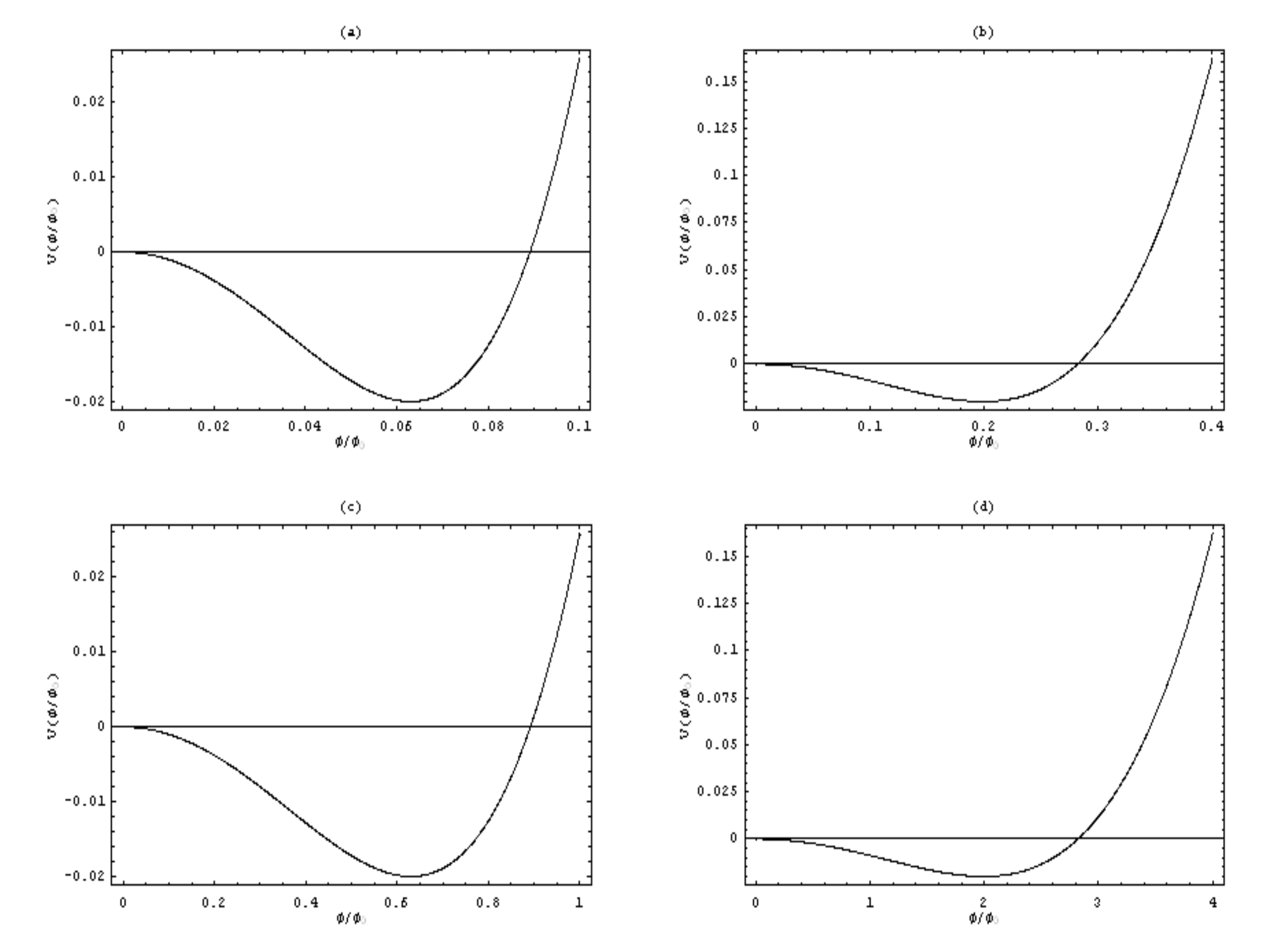}
\caption{In a) $T_c\sim T$ and $\rho=10\rho_0$, b) $T_c\sim T$ and $\rho=\rho_0$, c) $T_c\sim T$ and $\rho=0.1\rho_0$, and d) $T_c\sim T$ and $\rho=0.01\rho_0$.}
\label{fig:fig1_potencial}
\end{figure*}

Figure \ref{fig:fig1_potencial} shows $U(\phi/\phi_0)$. This potential is very smooth and therefore is expected that $\phi/\phi_0$ rapidly settles at $\phi_0$ after few oscillations and the particles produced by the parametric resonance mechanism ceases.

The minimum point as a function of $H_c$ decreases linearly to zero at the transition point. The transition in the potential from the superconducting state to the normal one near the transition point is given by \cite{landau2}
\begin{eqnarray}
\nonumber \Delta U(\phi)=\frac{VH_c^2}{8\pi},
\end{eqnarray}
\noindent where $V$ is the volume related with the mass of the superconducting field and its density through $V=n_{\pi}m_{\pi}/{\rho}$, where $n_{\pi}$ stands for the pion particle number present in $V$. Then
\begin{eqnarray}
\label{7} \Delta U(\phi)=\frac{n_{\pi}m_{\pi}H_c^2}{8\pi\rho},
\end{eqnarray}
\noindent and near the transition point this value corresponds exactly the energy gap due the transition from the normal state to the superconducting one. The friction term is comparable with $\phi_0$ and therefore can not be neglected.

The equation of motion to the superconducting pion field may finally written as
\begin{eqnarray}
\label{8}\ddot{\phi}+\frac{4\pi^2}{28\alpha\zeta(3)}\sqrt{\frac{\rho}{\pi}}H_c\dot{\phi}-H_c\sqrt{\frac{\rho}{\pi}}\phi+2\rho\phi^3=0.
\end{eqnarray}

\begin{figure*}[ht]
\centering
\includegraphics[width=14.0cm,height=10.0cm]{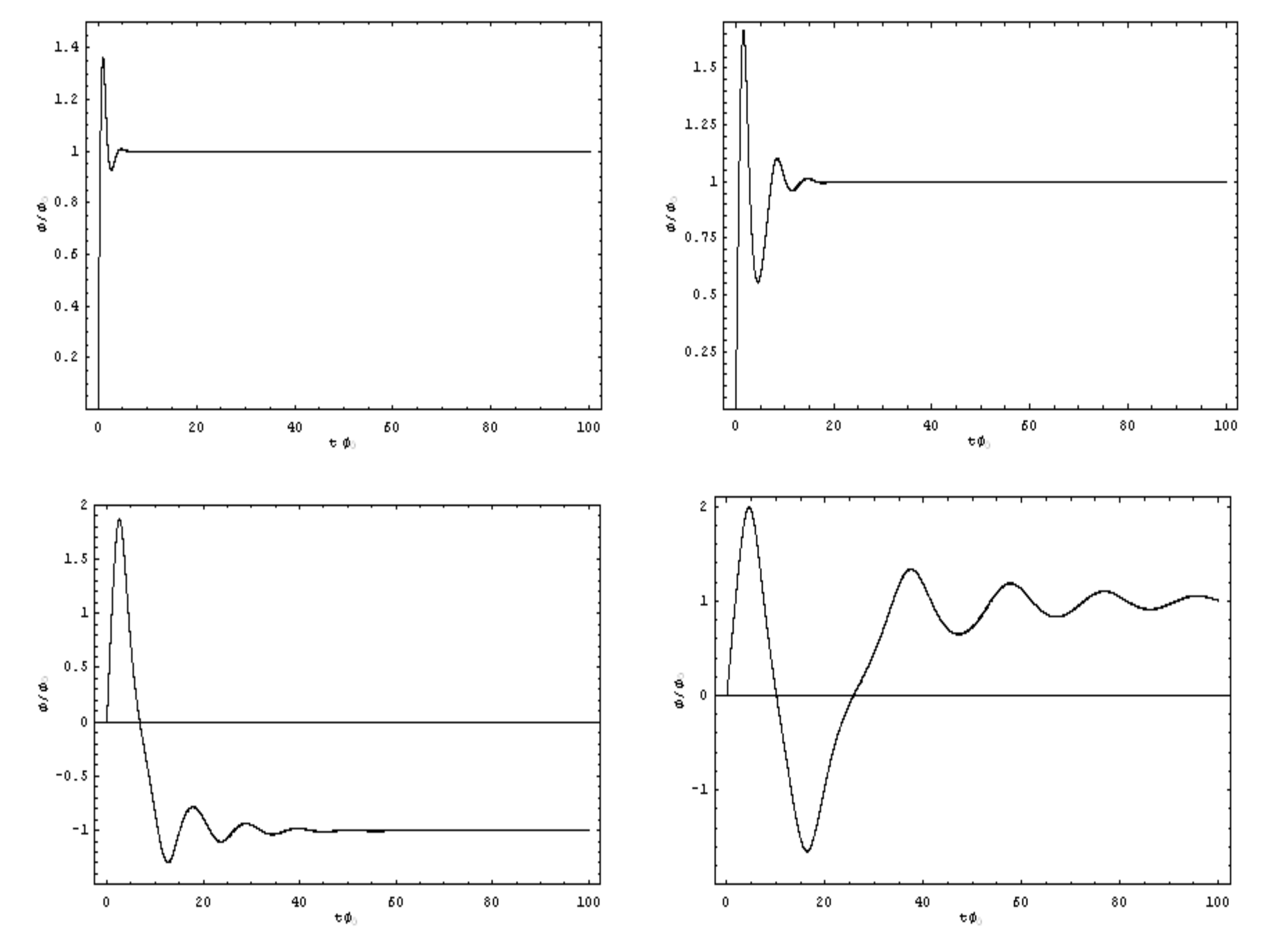}
\caption{The $\phi$ field as a function of $t\phi$. In a) $T_c\sim T$ and $\rho=10\rho_0$, b) $T_c\sim T$ and $\rho=\rho_0$, c) $T_c\sim T$ and $\rho=0.1\rho_0$, and d) $T_c\sim T$ and $\rho=0.01\rho_0$.}
\label{fig:fig2_phi}
\end{figure*}

This is basically a Duffing equation with a dumping term \cite{arscott} and it may be solved in terms of Jacobi elliptic functions \cite{erd}. Analytic solutions are a very hard matter and, otherwise, if one obtains analytical solutions then the parameters that arise from the different cases does not possesses an immediate physical meaning \cite{sdc} and the situation turns very complicated. Considering this, is performed here only a numerical study of the $\phi$ field evolution near the transition point. In Figure \ref{fig:fig2_phi} the behavior of $\phi/\phi_0$ as a function of $t\phi_0$ is showed. The number of oscillations is very small, i.e., the friction terms rapidly acts and therefore $\phi/\phi_0$ settles at $\phi_0$ and the process ceases. 

\section{Lagrangean Formalism}
A numerical analysis is performed using a Lagrangean toy model to study the particle production that may arise from the interaction between the superconducting field and the electromagnetic field of the neutron star. Such model, despite its simplicity, allows physical interpretation of the interaction process. Therefore, adopting this picture, the Lagrangean density may be written as 
\begin{eqnarray}
\label{x1}\mathcal{L}=-\frac{1}{4}F_{\nu\mu}F^{\mu\nu}+(2e)^2\psi^2A_{\mu}A^{\mu}+\mathcal{L_{\psi}},
\end{eqnarray} 
\noindent where $F^{\mu\nu}$ is the usual electromagnetic anti-symmetric tensor. Lagrangean density $\mathcal{L_{\phi}}$ contain terms that depends only on $\phi$ and therefore does not affect the equation of motion of the electromagnetic field. Applying the Euler-Lagrange equations to the electromagnetic field one obtains from (\ref{x1})
\begin{eqnarray}
\label{x2}\Box{A^{\mu}}-\partial^{\mu}\partial_{\nu}A^{\nu}+e^2(\pi^-\pi^+)A^{\mu}=0. 
\end{eqnarray}

Using a unitary gauge where the scalar field $\phi$ is real and the approach applied in \cite{sdc} one obtains the following equation of motion
\begin{eqnarray}
\label{final} \ddot{\chi}_k(t)+[k^2+2e^2\phi^2]\chi_k(t)=0,
\end{eqnarray}
\noindent in the momentum space. The back reaction of $\chi$ over the particles produced is a very strong component, i.e., the number of particles produced dominates the frequency of oscillations of the superconducting pion field turning its state. The back reaction may be computed as \cite{kofman}
\begin{eqnarray}
\label{backraction}\langle\langle \chi^2 \rangle\rangle=\frac{1}{2\pi^2}\int_0^{k_c}\frac{n_k(t)k^2}{k^2+2e^2\phi^2(t)}dk.
\end{eqnarray}

\begin{figure*}[ht]
\centering
\includegraphics[width=14.0cm,height=10.0cm]{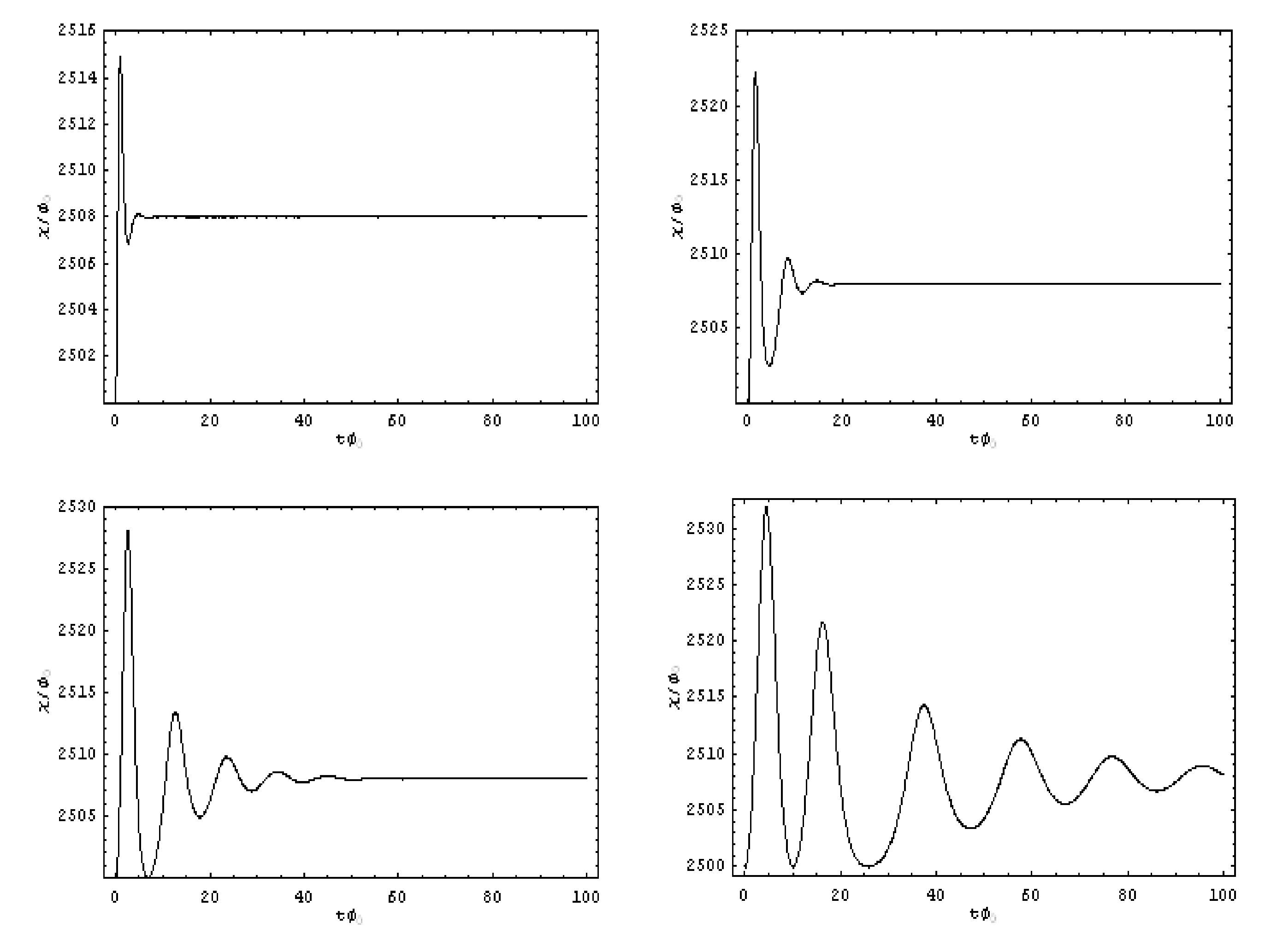}
\caption{Back reaction. In a) $T_c\sim T$ and $\rho=10\rho_0$, b) $T_c\sim T$ and $\rho=\rho_0$, c) $T_c\sim T$ and $\rho=0.1\rho_0$, and d) $T_c\sim T$ and $\rho=0.01\rho_0$.}
\label{fig:fig3_chi}
\end{figure*}

The superconducting field oscillations around the minimum of the effective potential rapidly vanish and the exponential term present in the above expression dominates, i.e., the back reaction avoid new particles production and the process ceases. In Figure \ref{fig:fig3_chi} is shown the back reaction process.

The equation (\ref{final}) is well-known and due to a Floquet theorem its solutions may be written as \cite{magnus}
\begin{eqnarray}
\label{x5}\chi_k(t)=e^{\mu_k(T_c-T)t}p(t),
\end{eqnarray}
\noindent where $p(t)$ is a periodic function that possess the same period of $\chi_k(t)$ and $\mu_k$ is the so-called Floquet exponent.

On the other hand the number of particles with a given momentum $k$ may be written as
\begin{eqnarray}
\label{x4} n_k(t){\simeq}e^{\mu_k(T_c-T)t}.
\end{eqnarray}

In a first approximation, i.e., in the first resonant band, one consider $\chi_k{\simeq}e^{\mu_k(T_c-T)t}$ and therefore one obtains that Floquet exponent behaves as the natural logarithm of particles produced.

\begin{figure*}[ht]
\centering
\includegraphics[width=14.0cm,height=10.0cm]{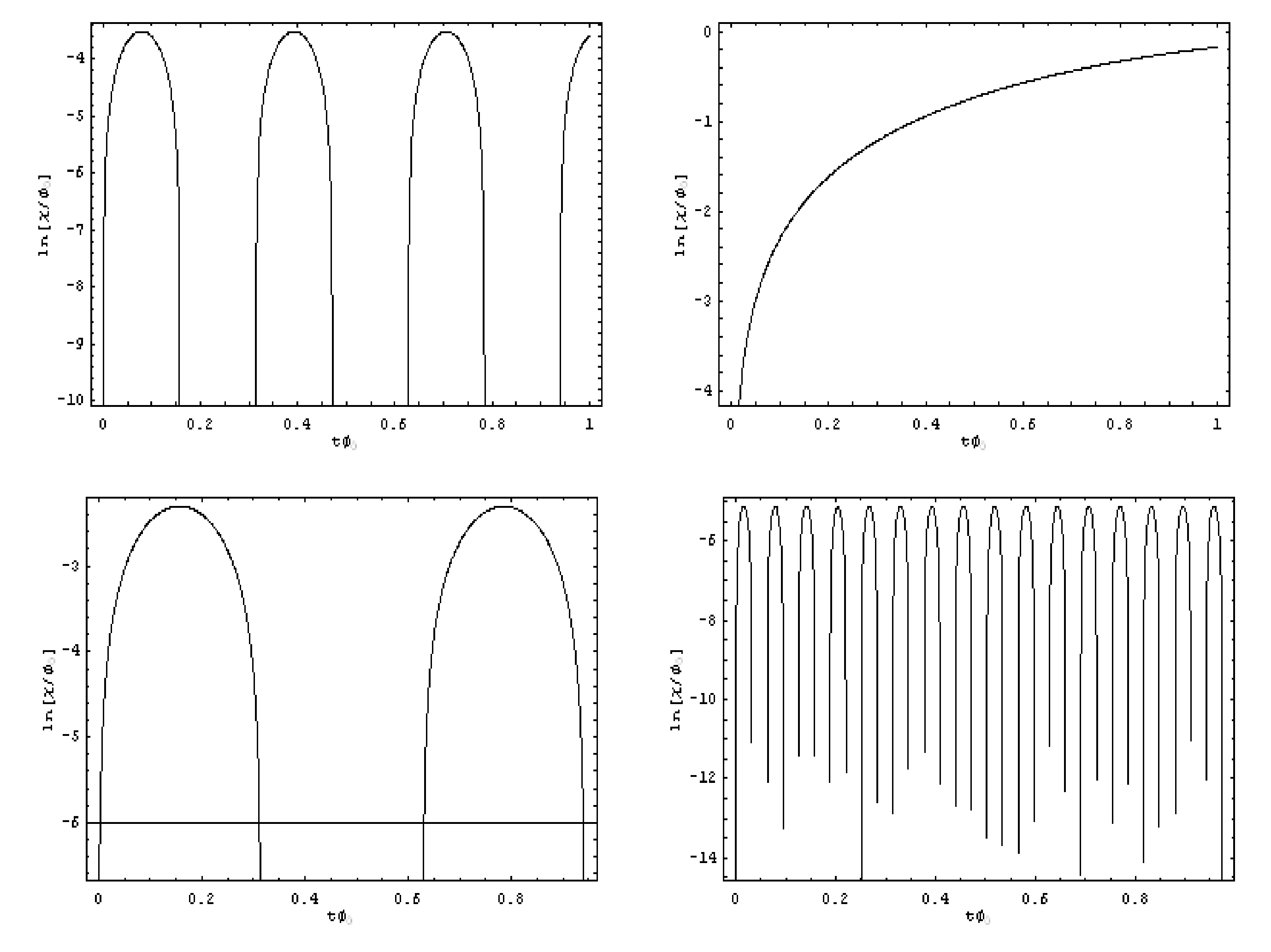}
\caption{Evolution of $\mu_k$ as a function of the logarithm of particles produced.}
\label{fig:f4_muk}
\end{figure*}

When $\mu_k$ is real the particles produced in the resonant band occur in an explosive way. Figure \ref{fig:f4_muk} show $\mu_kt$ $versus$ $ln[\chi_k(t)]$.

The energy density produced may be estimated by
\begin{eqnarray}
\label{x6}\rho_{\gamma}=\frac{1}{2\pi^2}{\int_0}^{k_c}\omega_k{n_k(t)}k^2dk,
\end{eqnarray}
\noindent where $k_c$ is the momentum cutoff. The above expression may be approximated in the first resonant band by
\begin{eqnarray}
\label{x7}\rho_{\gamma}\simeq\frac{1}{2\pi^2}{\int_0}^{k_c}\chi_kk^2dk\simeq\frac{1}{2\pi^2}\chi_kk^5.
\end{eqnarray}

\begin{figure*}[ht]
\centering
\includegraphics[width=9.0cm,height=7.0cm]{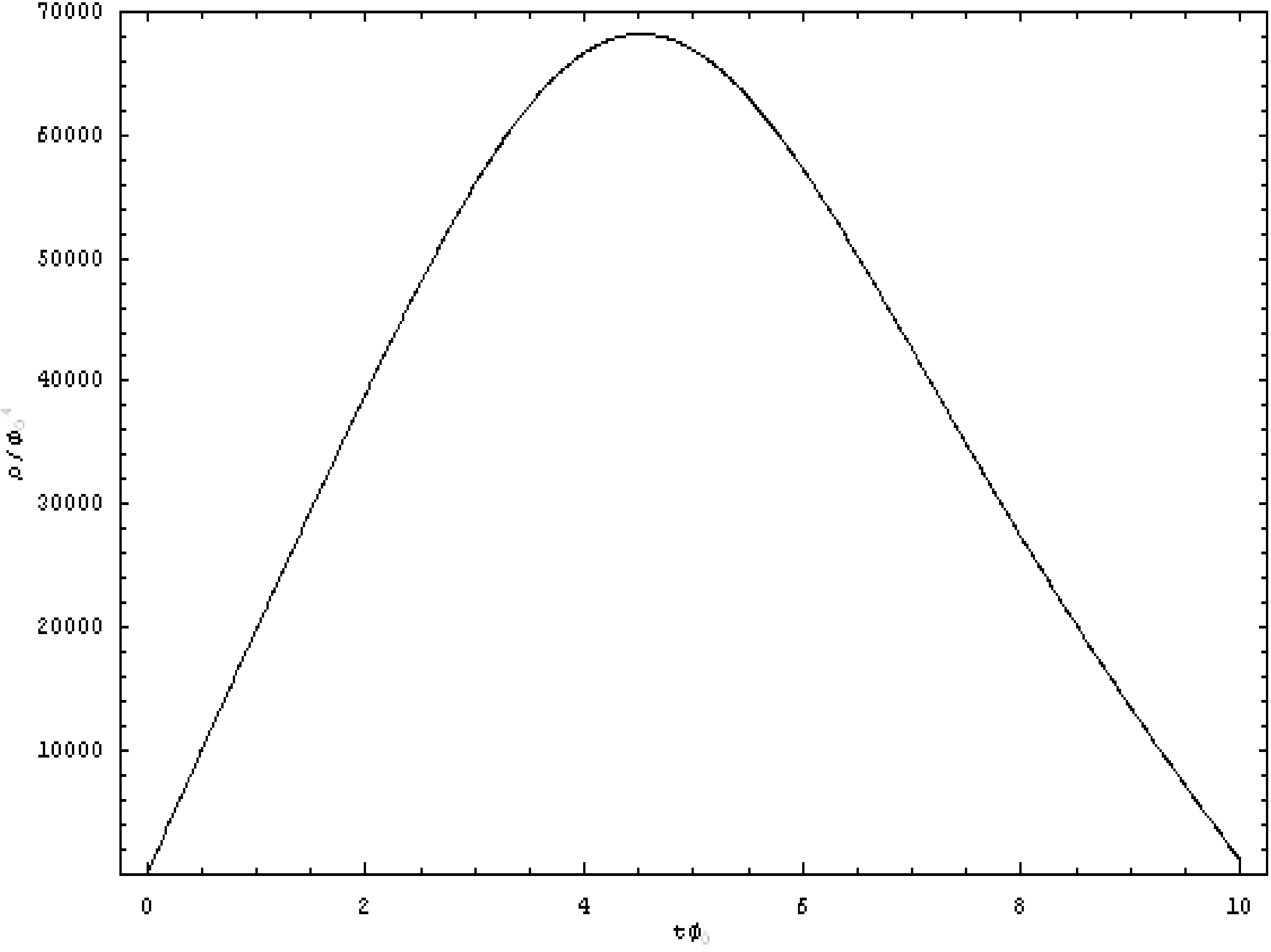}
\caption{Energy spectrum associated to the particle creation process in the first resonant band.}
\label{fig:f5_dens}
\end{figure*}

The energy density spectrum is similar to the black body radiation as shown in Figure \ref{fig:f5_dens}. 

\section{Final Remarks}
Parametric resonance may occur in neutron star medium due the interaction of a superconducting pion field and the electromagnetic field of the star. This particles production process is efficient, at least in the first resonant band, and allows one populate a region inside the star with a huge number of particles (photons). The photons produced will interact with the surrounding matter, increasing its temperature in a first moment. If the temperature growth is small, then the surrounding matter may absorbs completely the thermal energy and effects that might be observed at surface are negligible. Otherwise, if the growth of temperature are greater than the absorption capability of surrounding matter then the thermal energy excess coming from the particles produced may modify the surface temperature leading to a reheating in some region of the star outer crust. When the energy excess reach the star surface the outer crust may split arising plasma jets eventually observed in the Earth. 

The particle production is a non-equilibrium process and the temperature increase in the core may be calculated, for instance, using the Louville-von Newmann approach \cite{kim}. To take the correct increase of temperature at the star surface it is necessary to use a relativistic transport equation \cite{lindquist}. Modifications in the neutron star equation of state due to this particle production mechanism may be implemented allowing the study of new states of matter in extreme conditions of density and pressure. 

\section*{Acknowledgments}
The author is grateful to UFSCar for the financial support.


\end{document}